# Towards modelling cost and risks of infrequent events in the cargo screening process

*Ms. Galina Sherman*
*Prof. David Menachof*
University of Hull
Business School
Hull
G.Sherman@2008.hull.ac.uk, D.Menachof@hull.ac.uk

*Prof. Uwe Aickelin*
*Dr. Peer-Olaf Siebers*
University of Nottingham
School of Computer Science
Nottingham
uxa@cs.nott.ac.uk, pos@cs.nott.ac.uk

**ABSTRACT:**
*We introduce a simulation model of the port of Calais with a focus on the operation of immigration controls. Our aim is to compare the cost and benefits of different screening policies. Methodologically, we are trying to understand the limits of discrete event simulation of rare events. When will they become "too rare" for simulation to give meaningful results?*

Keywords: modelling and simulation, cost benefit analysis (CBA), cargo screening.

## 1. INTRODUCTION

The trade offs between the costs of risk policies and costs of the potential disruptions are very important in every business (Kleindorfer and Saad (2005)). In the context of port security, while protecting against the threats, the interest of the port authorities is to keep the port performance as smooth as possible. Thus CBA of different policies is important.

In port security, CBA is frequently discussed with regards to deciding policies for disastrous events, as for example terrorist attacks. However, as costs for such events often cannot be estimated, CBA analysis is limited (e.g. Bichou et al. (2009) and Jacobson et al. (2006)). A neglected application area for CBA is guiding policies for events that are relatively rare, but not disastrous, yet important to consider for planning investments in security, e.g. smuggling or illegal immigration. In this case, CBA promises to be more useful as compared to rare events where the cost often cannot be quantified. We shall call these situations "infrequent events".

In our research we want to show that CBA together with simulation is a useful tool for port authorities. Our hypothesis is that in addition to being more flexible and realistic than traditional CBA techniques such as decision trees, simulation will also allow us to better plan for infrequent events: Traditional CBA techniques are either based on mean outcomes or provide theoretic worst-case analysis. Simulation on the other hand allows us to look at a spread of outcomes over an appropriate number of replications, thus giving us a truer picture of the range of potential outcomes.

We believe that simulation is most useful if the probability of infrequent events is between certain upper and lower bounds. If the probability is too high, decision trees might be more efficient. If too low, too many replications will be required to be feasible. Establishing such bounds for port security will be one goal of our research.

In order to test this hypothesis, we conduct a case study at the ferry port of Calais where we look at illegal immigration attempts. We use simulation to estimate the likelihood of an illegal immigrant entering the country and CBA to evaluate different policies and estimate how much this cost the British taxpayer every year. Once our base model is in place we can then test different scenarios, to find out where to invest money.

This research is still at a very early stage. With regards to the simulation part of the project, we have collected knowledge and data about the port operation in Calais and we have built a first draft of the simulation model. With regards to CBA we have implemented a comparable model as a decision tree, a technology used in CBA.

In the remainder of the paper we will discuss about CBA and then present our case study site and operation in a bit more detail. Finally, we conclude and present our future work.

## 2. BACKGROUND

In security literature use of CBA is under discussion. Bichou et al. (2009) argue that conducting CBA while dealing with security issues will be impossible, because of the difficulty to estimate the costs of unwanted events. While Jacobson et al. (2006) propose using of CBA and simulation. In addition the authors suggest a cost benefit model that includes direct (e.g. operational costs) and indirect costs (e.g. false negatives). However, Jacobson et al. (2006) do not use the indirect cost component in their case study. Their argument is that while dealing with rare events it is very difficult to estimate the consequences.

Bjorkholm and Boeh (2006) suggest estimating the costs of cargo screening with x-ray machines and recommend better utilisation of this technology by using economic analysis. According to the authors there are three main reasons for screening: duties and taxes, illicit materials and external threats. The authors find that in first two cases economic analyses are possible while in the third case there is no agreement about the estimated costs.

Willis and La Tourrette (2008) use break even analysis in combination with limited simulation to estimate the benefits of a prevention of a terrorist attack. However, they agree that there are no complete reliable estimates. In addition the authors argue that using CBA for terrorism threats has multiple obstacles which sources are in: "uncertainty in terrorist risk level" and "very low recurrence interval for large attacks".

Farrow and Shapiro (2009) also suggest using break even analysis. According to the authors those analysis are implicit while CBA are explicit. Moreover, they argue that "with infrequent occurrences of some (not all) security events there will be an element of subjective or assumed probabilistic structure even in explicit models".

Another approach to ensure efficiency in the security investments is Markowitz theory of portfolio selection (1952). Talas and Menachof (2009) suggest using Markowitz's approach and build a conceptual model that helps to identify the efficiency of the security measures deployed.

Simulation has been used outside logistics to conduct CBA, e.g. in cancer screening research. Habbema et al. (1987), for example, use simulation and argue that CBA is an appropriate technique to compare two different cancer screening policies. The authors use a micro simulation approach for exploring different scenarios and their outputs. Similar to Habbema et al., Pilgrim et al. (2009) suggest conducting a cost effectiveness analysis for cancer screening policies while using discrete event simulation. These authors support their choice of methodology by previous research with the same research strategy.

## 3. CASE STUDY

In order to achieve our research as explained above, we have chosen to use a case study data collection approach. Calais has been chosen as our first case study for the following reasons: the number of links – Calais operates only with Dover; the simplicity of the cargo flow, only one particular threat of interest to the British government (clandestines). Clandestines are people who are trying to enter the UK illegally – without having proper papers and documents. In this case study the infrequent event is the probability to detect a lorry with additional freight on board. The cost of a clandestine to the UK per one year is £20,000 (UK Border Agency 2008).

Between April 2007 and April 2008 the throughput of lorries that passed screening facilities in Calais was greater than 900,000. Approximately, 0.4% of these lorries had additional human freight on board (UK Border Agency 2008). The number of undetected clandestines is unknown. Also very little is known about efficiency of the technology which is used. Some manufacturers have benchmarks but the benchmarks are according to an experimental environment rather than to a real world environment. It is very rare to find unbiased benchmarks that were made according to the real world.

The port is divided into two main parts: French and British control. Sensors are used on its perimeter: PMMW (passive millimetres waves) similar to x-ray machines; HB (heart beat detector) used to detect movements in the body of the lorry itself; $CO_2$ probe used to identify the level of oxygen inside the lorry; canine teams, and visual inspections. All the detectors listed above are in use by both French and British authorities, except for PMMW, which is in use only on the French side of Calais. Both authorities sort the lorries according to type (hard and soft sided). Generally, the PMMW detector is used for soft sided lorries, while HB detector is used for hard sided lorries. All other detection tools can be utilised for both types of the lorries.

For our case study we have built a decision tree that fully represents the flow inside our system. In addition we built a discrete event simulation model of the system by using AnyLogic software (XJ Technologies 2010), building both allows us to check the results and validate the model.

The basic simulation model represents the decision tree as closely as possible. However our simulation in comparison with the decision tree has an advantage of randomness (e.g. slightly different number of arrivals each time). We made ten replications of this basic model and calculated the average to compare its results with the results of the decision tree. In addition we changed the probability of a lorry being checked on the British side between 0% and 100% and compared the results with the results of the decision tree. We found that the simulation results (average of 10 replications) and the decision tree results are very similar, thus validating our simulation model.

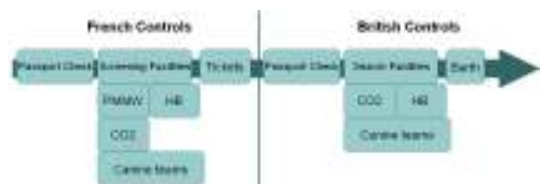

**Figure 1**: *The Conceptual Model of Calais Port*

## 4. CONCLUSIONS & FUTURE WORK

We think that CBA should not be ignored as a useful research tool in risk assessment. Currently we investigate different scenarios for our simulation model, e.g. exceeding queue capacities and adding peak and off peak times to the system. Following this, we will try to establish bounds for the probability of infrequent events versus the usefulness of simulation for CBA. Finally, we intend to apply our methodology to a second case study, the port of Dover.

**ACKNOWLEDGEMENT**

This project is supported by EPSRC, (EP/G004234/1) and the UK Border Agency.

**AUTHOR BIOGRAPHIES**

**GALINA SHERMAN** is a PhD student at Hull University, Logistics Institute. Her current research is related to supply chain management, risk analysis and rare events modelling.

**UWE AICKELIN** is Professor of Computer Science at The University of Nottingham, School of Computer Science. For more information see http://ima.ac.uk/aickelin

**DAVID MENACHOF** is Professor of Port Logistics at Hull Business School.

**PEER-OLAF SIEBERS** is a Research Fellow at The University of Nottingham, School of Computer Science. For more information see http://ima.ac.uk/siebers